\documentclass[letter,11pt]{article}

\usepackage{xcolor}
\usepackage{float}
\usepackage{graphics,graphicx}
\usepackage{hyperref,setspace}

\begin{document}

\begin{center}
\includegraphics[width=\textwidth]{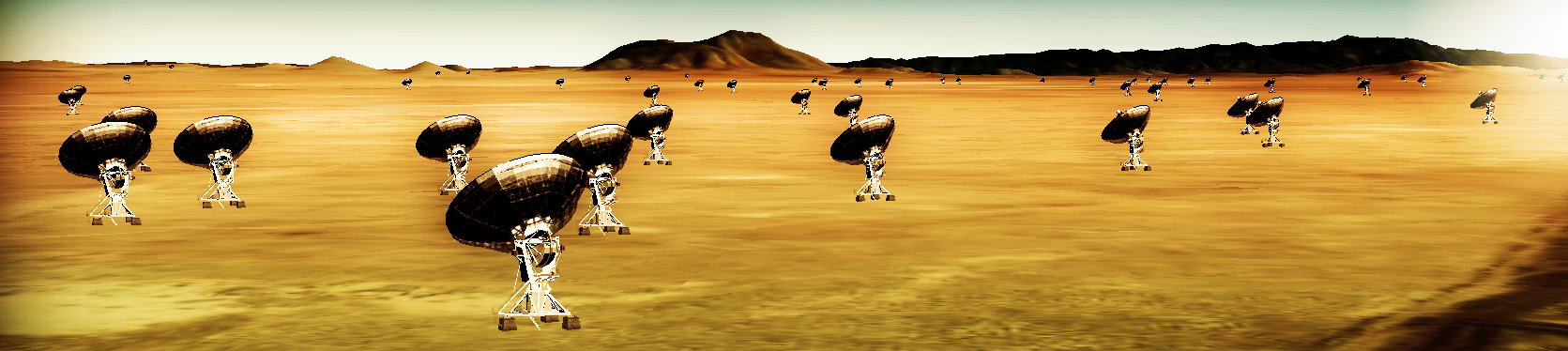}
\end{center}

\begin{center}

{\bf \large Next Generation Very Large Array Memo No. 103}

\vspace{0.1in}

{\bf \large Spectroscopy of High Redshift Galaxies with the ngVLA}

\end{center}

\hrule 

\vspace{0.3cm}

\centerline{C.L. Carilli}
\centerline{National Radio Astronomy Observatory, Socorro, NM, USA}

\vspace{0.1in}

\centerline{Marcel Neeleman}
\centerline{National Radio Astronomy Observatory, Charlottesville, VA, USA}

\vspace{0.3cm}

%\clearpage
%\newpage

\begin{abstract}

We present simulations of the capabilities of the ngVLA to image at $\sim 0.75$ kpc resolution ($0.085"$), molecular line emission from star forming disk galaxies at high redshift. The results are compared to the current capabilities of ALMA. ALMA can detect the integrated emission, and determine the velocity gradient and size across the brighter emission regions of the galaxy. The ngVLA is a factor $\sim 6$ more sensitive at the adopted spatial and velocity resolution. This sensitivity is needed to recover the detailed column density distribution, velocity field, and velocity dispersion at full resolution. The ngVLA will enable detailed analysis of spectral line profiles at $0.75$~kpc resolution, even in relatively faint regions. The ngVLA will trace the rotation curves to large radii, and recover sub-structure in the disks, such as clumps, spiral arms, bars, and rings. Detection of these features is crucial in order to assess how cold gas precipitates the formation of stars at high redshift.

\end{abstract}

\section{Introduction}

A number of memos have investigated the capabilities of the ngVLA to image the molecular and other redshifted millimeter line emission from high redshift galaxies (ngVLA memos 8, 13, 15, 50\footnote{https://ngvla.nrao.edu/page/memos}). In this work, we explore this important science case (KSG 3\footnote{https://ngvla.nrao.edu/page/scibook}) in more detail, incorporating a realistic model based on the spatial and dynamical properties of a nearby galaxy, and employing the latest ngVLA configuration. We compare the results to the capabilities of the current ALMA.

The principle application will be in the imaging of CO line emission from intermediate to high redshift galaxies at $0.085" \sim 0.75$kpc (Carilli and Walter 2013). However, the results are generally applicable to any line emission (or absorption), from high redshift galaxies of similar line strength and spatial distribution. Possibilities include: (i) [CI] 492 GHz at $z \sim 4$ to 5 (Valentino et al. 2018), (ii) H$_2$O absorption against the CMB at $z \sim 4$ to 7 (Riechers et al. 2022), and (iii) extreme redshift ($z \sim 20$) fine structure line emission (Carilli et al. 2017). 

\section{Model}

As an input model, we adopt the HI 21cm emission cube from the THINGS survey of the galaxy NGC 2366 (Walter et al. 2008). The galaxy is classified as a barred irregular dwarf galaxy. This galaxy shows a rotating HI disk that is modestly disturbed, with a possible outer warp (van Eymeren et al. 2009). The integrated line profile is flat-top, not a clear double horn. Such a profile is more characteristic of those seen for CO emission from high redshift disk galaxies, for which classic double horn line profiles are uncommon (Gonzales-Lopez et al. 2019). The THINGS survey has a resolution of a few hundred parsecs, and provides reasonably high HI 21cm signal-to-noise spectra across the disk, on scales out to twice the radius of the main optical galaxy in the case of NGC 2366. 

The importance of the input model is simply to provide a dynamical template that could be representative of a real galaxy at high redshift. The strength, size, and full velocity width then have to be adjusted to match those expected for a disk galaxies at high redshift. We adopt parameters that are representative of the massive main sequence star forming disk galaxies seen in ASPECS -- the deepest blind, large cosmic volume search for molecular gas to date (Walter et al. 2016): 

\begin{itemize}
    \item Size: full extent = 0.8" ($\sim 7$ kpc)
    \item Line FWHM = 400 km/s 
    \item Line peak = 1.2 mJy
    \item Integrated flux = 414 mJy km/s (derived from Gaussian fit)
    \item Central frequency = 86 GHz (eg. CO 2-1 at $z = 1.7$, or 3-2 at $z = 3.0$)
\end{itemize}

As an example, if this were CO 2-1 emission at $z = 1.7$, the molecular gas mass would be $1.6\times 10^{11}$ M$_\odot$ for Milky Way excitation and gas mass-to-line luminosity conversion, or 
$1.9\times 10^{10}$ M$_\odot$ for starburst galaxy conditions (Carilli \& Walter 2013). 

The line strength, source size, and line width are also comparable to what was seen in the recent discovery of H$_2$O absorption against the CMB at $z = 6.3$ by Riechers et al. (2022). Hence, the results of these simulated observations can be used as a guide to the capabilities of the ngVLA to provide spatially resolved observations of this new cosmic phenomenon -- a phenomenon that can be used to determine the evolution of the temperature of the CMB to very high redshift. Multiple lines of sight through a given source may prove crucial in mitigating systematic uncertainties due to source line and continuum spatial structure. 

On a more speculative note, the 86 GHz model corresponds to [CII] 158 $\mu$m line at $z = 21$, with a line luminosity of $2.6\times 10^9$ L$_\odot$. Such a [CII] luminosity is at the high end of the distribution of [CII] luminosities for galaxies discovered in the ALPINE and REBELS surveys at $z \sim 4$ to 7 (Yan et al. 2020; Bouwens et al. 2022). Whether such galaxies exist at $z = 21$ remains unknown.

The input image cube is blanked at 3$\sigma$ in each channel to remove noise. The frequency structure of the cube is then set to a channel resolution of 2.5 MHz = 8.7 km s$^{-1}$, with a total of 66 channels. 

\section{Simulations and Imaging}

SIMOBSERVE in CASA was used for the simulations. We simulate a four hour synthesis with 20 second records, and scale the noise for a total integration time of 20 hours (ie. 5 observations). 

We employ the ngVLA Rev D configuration, including the Core and Spiral components (ngVLA memo 92). For ALMA, we use the c9 configuration (longest baseline of 13.6 km). This ALMA configuration results in a natural weighted synthesized beam FWHM of $0.091"\times 0.077"$.

TCLEAN was used to image and deconvolve the spectral cubes. A channel width of 12.5 MHz (44 km s$^{-1}$) was used for the image cubes (width = 5 relative to the input measurement set). The cubes were cleaned to a residual of about $2\sigma$. For ALMA we used natural weighting, leading to the synthesized beam given above, and an rms noise of 54 $\mu$Jy channel$^{-1}$. 

For the ngVLA, we used Briggs weighting with R=0 and a Gaussian taper of $0.081"$. This leads to a beam FWHM of $0.086"\times 0.083"$, and an rms noise of 9.0 $\mu$Jy channel$^{-1}$. This noise is a factor 1.5 larger than the noise expected for the Main array (214 antennas)  using natural weighting.\footnote{The declination of the model had to be adjusted to $-30^o$ for ALMA, and $+30^o$ for the ngVLA.}

\section{Analysis}

\subsection{Spectra and Moment Maps}

Figure 1 shows the results for the integrated spectrum of our target galaxy by integrating all flux within $0.5''$ radius of the center of the galaxy. Shown in this, and subsequent, figures with spectra are: 

\begin{itemize}

    \item The ngVLA simulated observation spectrum at 44 km s$^{-1}$ channel$^{-1}$, and noise appropriate for a 20 hour observation.

    \item The ALMA simulated spectrum at 44 km s$^{-1}$ channel$^{-1}$ with noise appropriate for a 20 hour observation.
    
    \item The Gaussian fit to the ngVLA noisy data.
    
    \item The input model as imaged with the ngVLA but with no noise added, and at 8.7 km s$^{-1}$ channel$^{-1}$. We call this the 'input model', since it represents effectively the input model smoothed to the target spatial resolution of the observations ($\sim 0.085"$). Thus representing the 'best' one can do in the absence of noise at this spatial resolution.

\end{itemize}

\begin{figure}
\centering 
%\hspace*{-3cm}
\centerline{\includegraphics[scale=0.4]{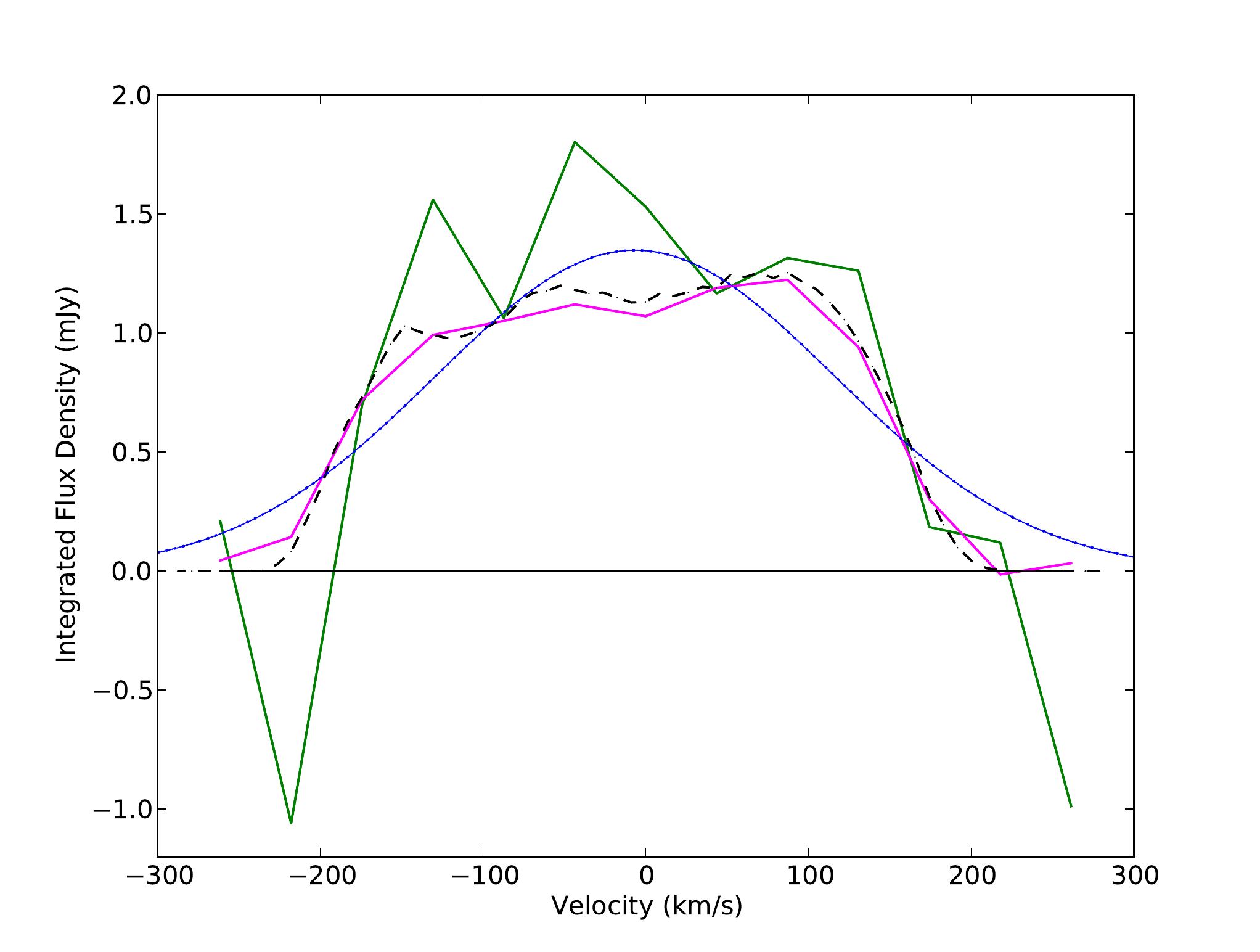}}
\caption{Integrated spectrum of the model source. The dash line is the input model at 8.7 km s$^{-1}$ channel$^{-1}$. The magenta curve is the ngVLA measurement at 44 km s$^{-1}$ channel$^{-1}$. The blue dotted curve is a Gaussian fit to the ngVLA data. The green curve is the ALMA measurement at 44 km s$^{-1}$ channel$^{-1}$.
}
\end{figure}  

Table 1 shows the results for Gaussian fitting to these four spectra. The model row is roughly 'truth' (as far as a Gaussian can parameterize the profile), since no noise is added, and the input velocity resolution is much higher. 

For the integrated spectrum, the ngVLA results are within a few percent of the correct line peak, mean velocity, velocity width, and velocity integrated flux. The ALMA results deviate from the model by about 20\% to 30\%. 

\begin{table}
\centering
\footnotesize
\caption{Line Gaussian Fitting Results}
\begin{tabular}{lcccc} 
  \hline
  \hline
  Spectrum & Peak & Velocity & Dispersion & Area  \\
  ~ & mJy & km s$^{-1}$ & km s$^{-1}$ & mJy km s$^{-1}$ \\
  \hline
Integrated & ~ & ~ & ~ & ~ \\
Model & $1.34\pm 0.04$ & $-6.5\pm 5$ & $122\pm 5$ & $414\pm 21$ \\
ngVLA & $1.30\pm 0.11$ & $-7.6\pm 12$ & $125\pm 13$ & $ 404\pm 52$ \\
ALMA & $1.72\pm 0.38$ & $-9.4\pm 26$ & $103\pm 61$ & $446\pm 150$ \\
 \hline
254, 258 & ~ & ~ & ~ & ~ \\
Model & $0.205\pm 0.002$ &  $58.4\pm 0.7$ & $54.4\pm 0.7$ & $28.1\pm 0.5$ \\
ngVLA & $0.216\pm 0.01$ & $56.8\pm 2.6$ & $53.3\pm 2.6$ & $29.0 \pm 1.8$ \\
ALMA & $0.16\pm 0.04$ & $24.6\pm 17$ & $61.8\pm 17$ & $24.8\pm 9$ \\
 \hline
 259, 250 & ~ & ~ & ~ & ~ \\
Model & $0.209\pm 0.002$ & $-88.3\pm 0.7$ & $53.1\pm 0.7$ & $27.8 \pm 0.5$ \\
ngVLA &  $0.204 \pm 0.009$ & $-81.1\pm 3.0$ & $55.4\pm 3.0$ &  $28.4 \pm 2.1$ \\
ALMA & $0.33\pm 0.033$ & $-74.1\pm 5.3$ & $46.3\pm 5.4$ & $38.2\pm 6$ \\
 \hline
250, 266 & ~ & ~ & ~ & ~ \\
Model & $0.163\pm 0.002$ & $127.0\pm 0.7$ & $41.1 \pm 0.7$ & $16.8\pm 0.2$ \\
ngVLA &  $0.160\pm 0.012$ & $125.2\pm 3.7$ & $42.9\pm 3.7$ & $17.2\pm 2.0$ \\ 
ALMA & ~ & ~ & ~ & ~ \\
 \hline
 264, 237 & ~ & ~ & ~ & ~ \\
Model & $0.074\pm 0.001$ & $-165.7\pm 0.3$ & $25.0\pm 0.3$ & $4.64\pm 0.07$  \\
ngVLA & $0.064\pm 0.014$ & $-157.7\pm 7.8$ & $32.8\pm 9.1$ & $5.3\pm 2.0$ \\
ALMA & ~ & ~ & ~ & ~ \\
 \hline
\hline
\vspace{0.1cm}
\end{tabular}
\end{table}

Figures 2, 3, and 4 show the velocity-integrated line intensity, mean velocity and velocity dispersion images created with the python package Qubefit (Neeleman et al. 2021). For the velocity integrated line intensity we integrate the line emission between $-225$ and $+175$ km s$^{-1}$. The mean velocity and velocity dispersion field are the peak and standard deviation of a Gaussian fit of the data cube at the given position. This approach is more robust than the standard approach of $\sigma$-clipping and calculating moments for the resolution and signal-to-noise considered here (see Neeleman et al. 2021), with the caveat that the emission line profile is assumed to be Gaussian everywhere.

The ngVLA velocity integrated line intensity, recovers source structure, including the small scale structure at higher brightness, and the more diffuse extended structure out to $0.4"$ radius. ALMA recovers the brighter peaks in the inner $0.2"$ radius, and little else.

Similar can be said for the mean velocity field and the velocity dispersion field: the ngVLA recovers velocity structure across the disk, while ALMA shows evidence for a velocity gradient only in the inner disk.

Figure 5 shows point spectra at full spatial resolution for four positions in the disk: two regions of high surface brightness (peak $\sim 0.2$ mJy beam$^{-1}$), one of intermediate brightness ($\sim 0.15$ mJy beam$^{-1}$), and one of lower surface brightness ($\sim 0.07$ mJy beam$^{-1}$). 

The Gaussian fitting results for these point spectra (Table 1) indicate that, for the bright regions, both ALMA and ngVLA recover reasonable estimates of the line parameters, with the ngVLA providing a few percent accuracy results, and ALMA more like 30\% accuracy. 

For the fainter two regions, Gaussian fitting to the ALMA spectra fail, while the ngVLA still provides few to 10\% accuracy for the parameters.

\begin{figure}
%\centerline{\includegraphics[scale=0.75]{Mom0.pdf}}
\includegraphics[width=\textwidth]{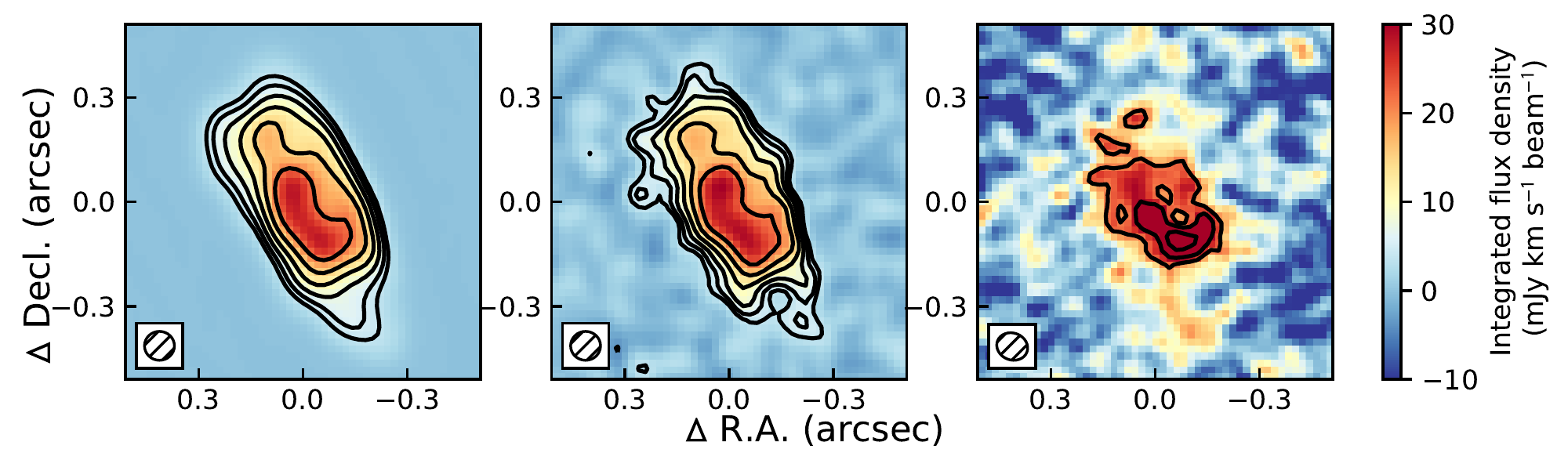}
\caption{Velocity integrated emission. Left: input model (in this, and subsequent, images, the input model is convolved to the target resolution). Center: ngVLA at $0.085''$ resolution in 20hrs. Right: ALMA at $0.085''$ resolution in 20hrs. In the ngVLA and ALMA images, the contour levels start at 3$\sigma$ and increase by powers of $\sqrt{2}$, where $\sigma$ is 1.3 and 7.0 mJy km s$^{-1}$ beam$^{-1}$ for the ngVLA and ALMA image respectively. The contours of the input model are drawn at the same level as the ngVLA image. The inset shows the beam size of the images.
\label{fig:moment0}}
\end{figure}  

\begin{figure}
%\centerline{\includegraphics[scale=0.75]{Mom1.pdf}}
\includegraphics[width=\textwidth]{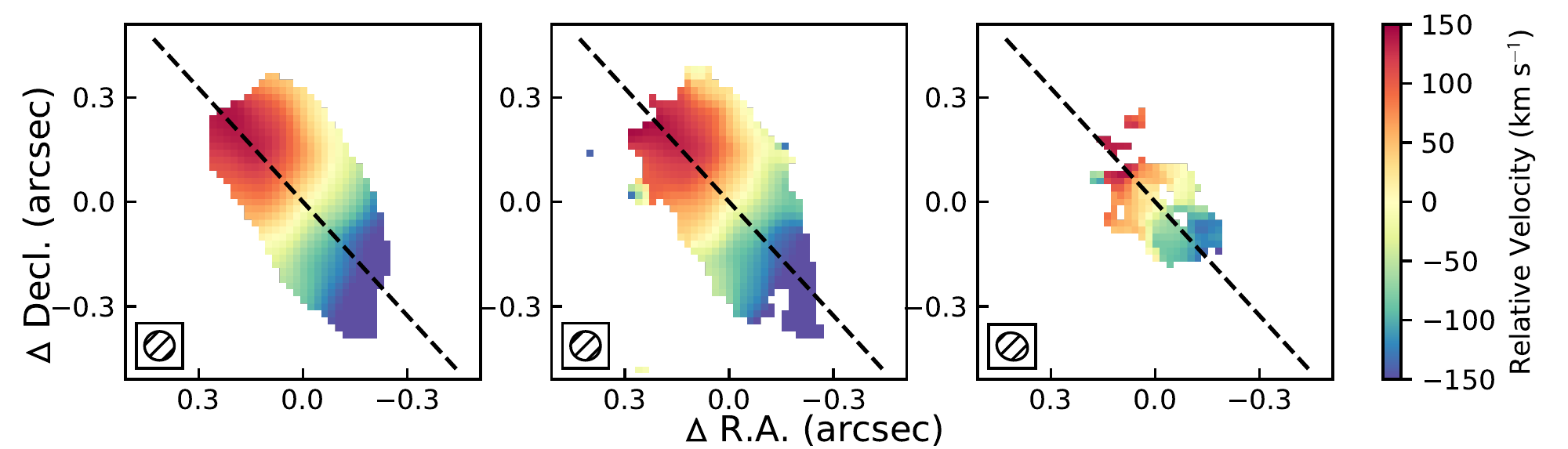}
\caption{Mean velocity field. Left: input model. Center: ngVLA at $0.085''$ resolution in 20hrs. Right: ALMA at $0.085''$ resolution in 20hrs. The dashed black line is the major axis long which the position-velocity diagram of Figure \ref{fig:pv} are generated.
\label{fig:moment1}}
\end{figure}  

\begin{figure}
%\centerline{\includegraphics[scale=0.8]{Mom2.pdf}}
\includegraphics[width=\textwidth]{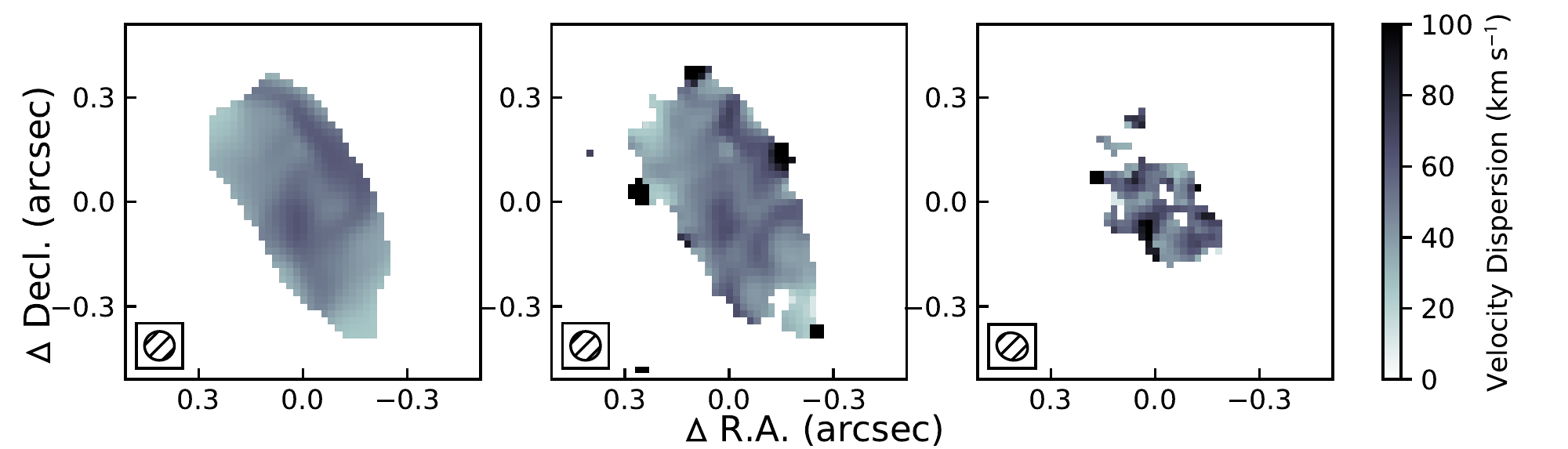}
\caption{Velocity dispersion field. Left: input model. Center: ngVLA at $0.085''$ resolution in 20hrs. Right: ALMA at $0.085''$ resolution in 20hrs. Note the black regions at the edge of the galaxy are regions where the velocity dispersion is hard to measure and have subsequently larger uncertainty associated with it.
\label{fig:moment2}}
\end{figure}

\begin{figure}
\centering 
%\hspace*{-3cm}
\centerline{\includegraphics[scale=0.7]{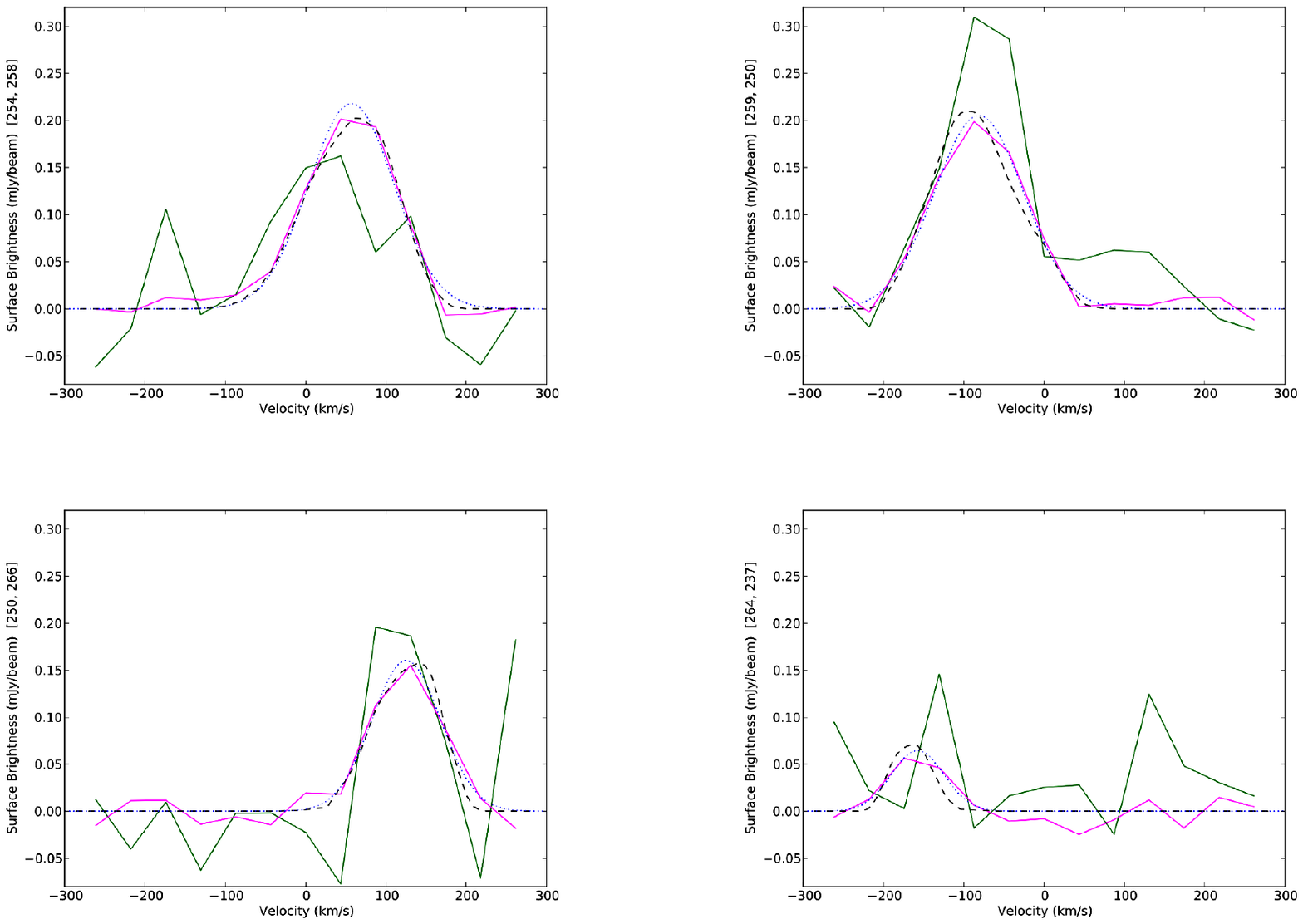}}
\caption{Spectra at four positions at $0.085''$ resolution. The dash line is the input model at 8.7 km s$^{-1}$ channel$^{-1}$. The magenta curve is the ngVLA measurement at 44 km s$^{-1}$ channel$^{-1}$. The blue dotted curve is a Gaussian fit to the ngVLA data. The green curve is the ALMA measurement at 44 km s$^{-1}$ channel$^{-1}$
}
\end{figure}  

\subsection{Rotation Curve Analysis}

\begin{figure}
\centering 
\includegraphics[width=\textwidth]{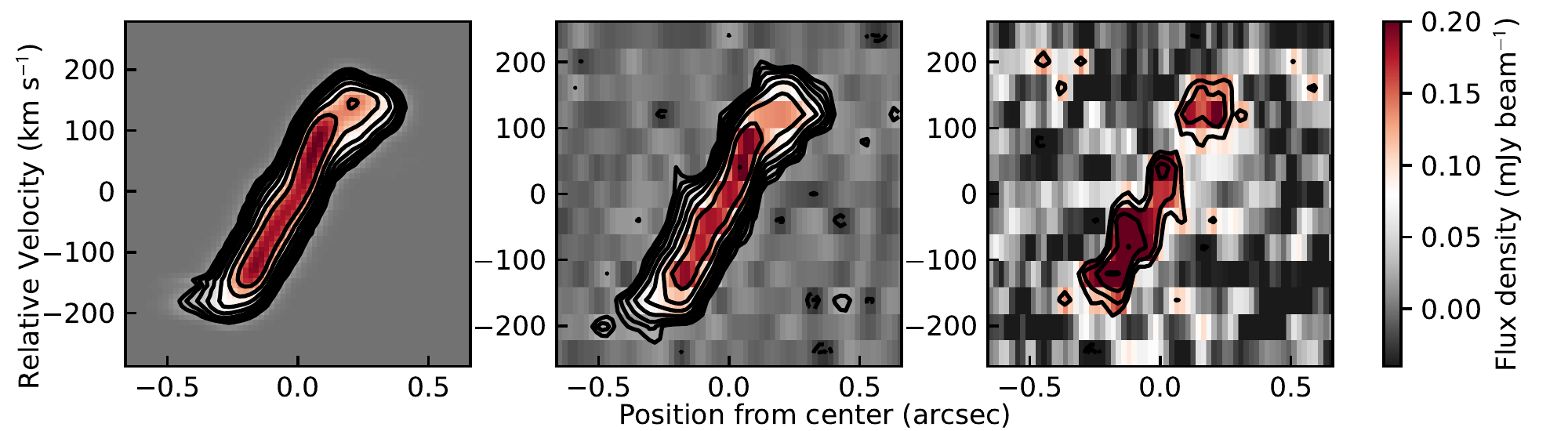}
\caption{Position--velocity diagram along the major axis of the galaxy. The left panel is the input model with a channel spacing of 8.7 km s$^{-1}$. The middle and right panel are the ngVLA and ALMA observations at a channel spacing of 44 km s$^{-1}$, respectively. For the ngVLA and ALMA imaging, the contours start at twice the rms noise of the data cube and increase in powers of $\sqrt{2}$, for the model image the contours are drawn at the same level as the ngVLA contours. The center of the x-axis is the kinematic center of the galaxy. 
\label{fig:pv}}
\end{figure}

In Figure \ref{fig:pv}, we show the position--velocity (p--v) diagram of the data cubes taken along the major axis of the galaxy using the python package pvextractor (Ginsburg et al. 2015). These p--v diagrams are used to assess the rotation curve of the galaxy. For the model input (left panel), we can see several features that are worth highlighting. First is the velocity gradient along the major axis, which is a characteristic of rotating disk galaxies. This velocity gradient is well-produced by both the ALMA and ngVLA observations in magnitude, but the ngVLA observations recover the full extent of the emission, whereas ALMA only recovers the inner part of the structure. Recovering this extended emission allows for a better characterization of the rotation curve at large galactocentric distances, where the effect of dark matter starts to dominate the contribution to the rotation curve (e.g., Genzel et al. 2020). 

Second, the model galaxy, NGC 2366 has several departures from pure disk rotation, as can be seen in Figure \ref{fig:moment1} and is discussed in Hunter et al. (2001). Two such features are the kink in the velocity gradient between the northern part and southern part of the galaxy as well as the extra emission seen in the northern part of the galaxy at positive velocities. Both these features are clearly visible in the ngVLA observations but are not detectable in the ALMA observations. Deviations from a regular rotating disk model are important to assess corrections on the rotation curve and to classify kinematic effects due to outside torques induced by mergers or the galaxy's halo (e.g., J\'{o}zsa 2007).

\subsection{Galaxy ISM Structure}

\begin{figure}
\centering 
\includegraphics[width=\textwidth]{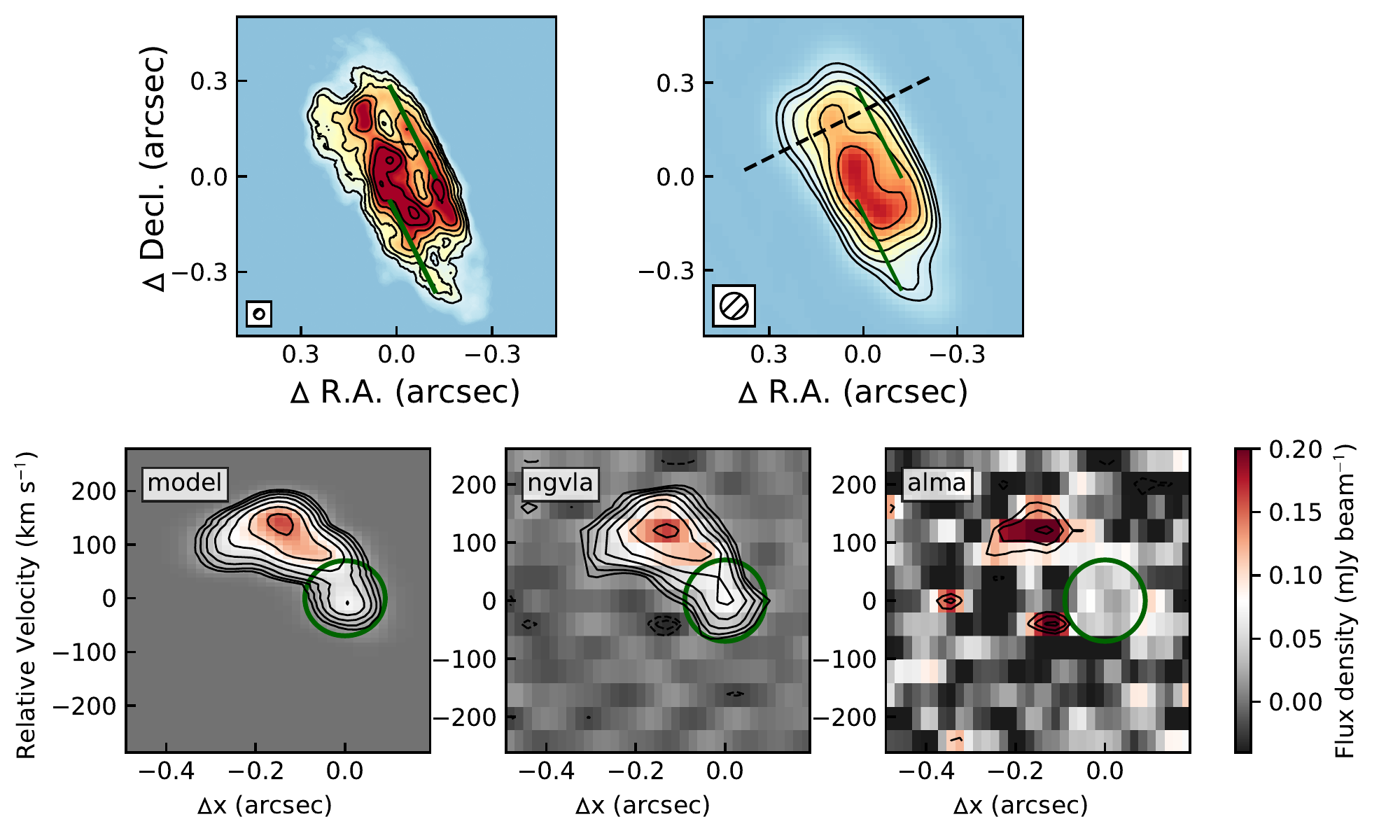}
\caption{Position--velocity diagrams for the ridge feature seen in the galaxy. The input galaxy model has two distinct ridges running parallel to the major axis (green solid lines in top figures), which can be seen at the resolution of the original data in the top left panel and convolved to the observed resolution in top right panel. Position--velocity diagrams along the dashed black line, which is perpendicular to the ridge (and major axis), are shown in the panels below. The ridge can be observed as excess emission at the origin (green circle). Contours are as in Figure \ref{fig:pv}. 
\label{fig:ridge}}
\end{figure}

\begin{figure}
\centering 
\includegraphics[width=\textwidth]{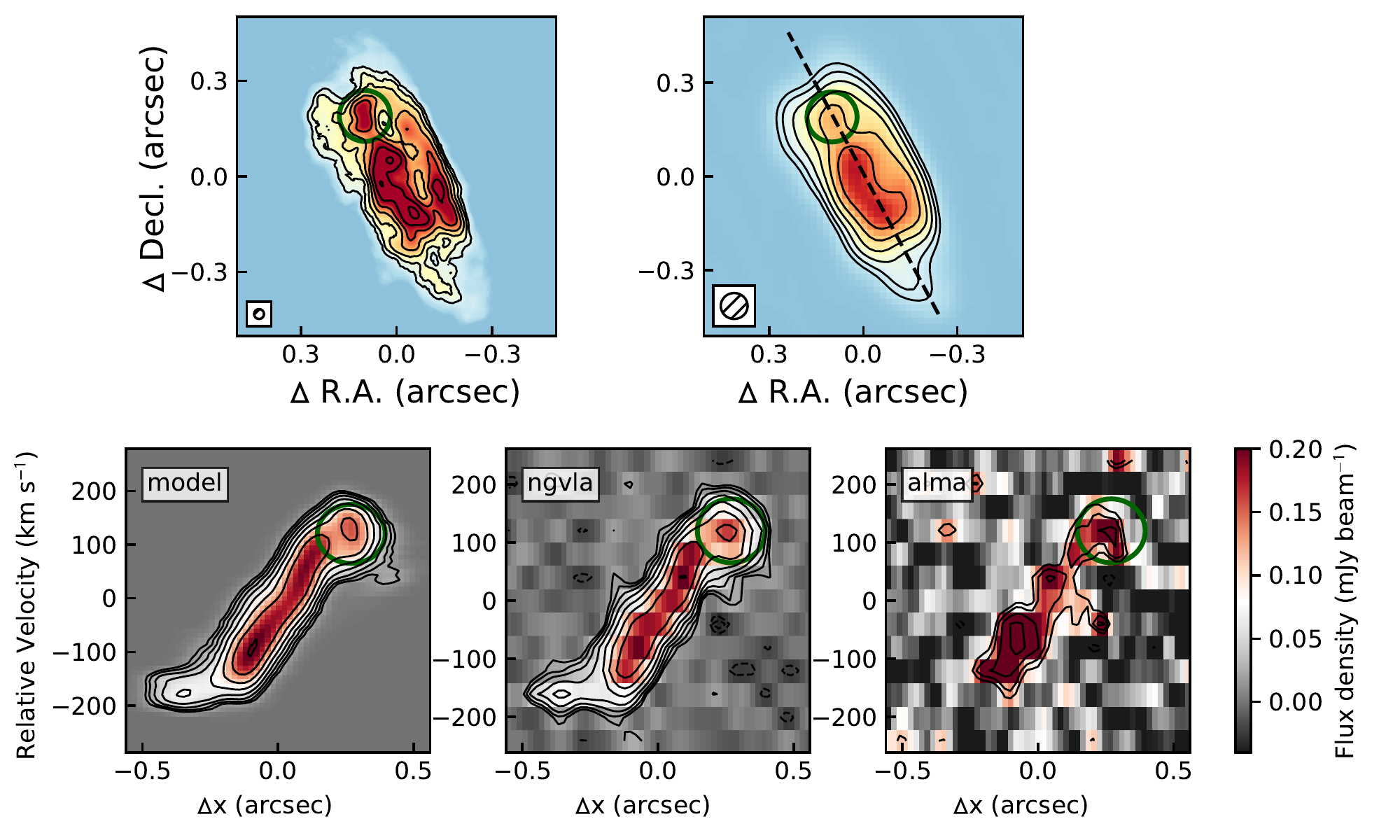}
\caption{Same as Figure \ref{fig:ridge} except for the emission clump (green circle). 
\label{fig:clump}}
\end{figure}

Our input galaxy model, NGC 2366, has several ISM features that would be interesting to explore at high redshift. In particular, this galaxy contains two HI ridges that run parallel to the major axis, possibly forming an HI ring (Hunter et al. 2001) or weak spiral arms (Van Eymeren et al. 2008). In addition, there are several HI clumps throughout the disk of the galaxy (Hunter et al. 2001). The HI ridges and one of the HI clumps are shown in the top left panels of figures \ref{fig:ridge} and \ref{fig:clump} at the resolution of the HI data. In the top right panel, we show the same data convolved to the resolution obtainable with the ngVLA (and ALMA) for a high redshift galaxy. This figure, and the accompanying position-velocity diagram (bottom left panels), show that both the HI ridge and clump can be resolved at this resolution. In the position-velocity diagram, the HI ridge shows up as excess emission at 0 km s$^{-1}$ (green circle) due to the HI ridge having a lower rotation speed than the surrounding gas (this is what is causing the kinematic warp seen in Figure \ref{fig:moment1}). For the HI clump, the position-velocity diagram shows excess emission in-line with the rotation curve of the galaxy. Both features are therefore clearly distinguishable at this resolution provided that there is enough sensitivity in the observations.

To assess the sensitivity that is needed to recover these ISM features, we explore the simulated data sets of both the 20 hour ngVLA observation and the 20h ALMA observation in configuration C-9. The corresponding position-velocity diagrams are shown in the bottom middle and left panel for the ngVLA and ALMA observations, respectively. These panels show that the 20 hour observations are sufficient to both recover the emission in the ridge as well as the clump at $>$8 $\sigma$. This implies that these features will be detected at $>$4 $\sigma$ in just 5 hours of ngVLA observing time. This is in contrast with the 20 hour ALMA observations. At the position of the ridge, no significant emission is detected in the ALMA observations (bottom right panel), and although there is emission detected at the position of the clump, the sensitivity of the observations is insufficient to discern if this is a true clump or a noise fluctuation. 

The ngVLA will therefore provide a unique view of the density variations within the ISM of high redshift galaxies. Detecting such density variations at `cosmic noon' is crucial in order to understand how ISM variations precipitate star formation as measured via optical tracers such as H$\alpha$ (e.g., Erb 2006).

\section{Discussion}

We find that at 86~GHz, a target resolution of $0.085"$ (0.75~kpc), and fixed velocity resolution, the ngVLA obtains a sensitivity a factor six better than the optimal ALMA configuration. For observations of a representative high redshift line emitting source, such as CO from a main sequence disk galaxy, in 20 hours ALMA provides a reasonable integrated line profile (parameters accurate to $\sim 20$ to 30\%), a lower limit to the source size, missing the fainter outer disk, and a determination of the velocity gradient for the brighter inner disk. In 20 hours, the ngVLA obtains an accurate integrated spectrum (few percent in profile parameters), and accurate moment images (column density, mean velocity, dispersion), across the full disk. 

In terms of point spectra at full resolution, ALMA again obtains modest accuracy line profile parameters ($\sim 20$ to 30\%) in the brighter regions of the source, but fails in the fainter, outer disk. The ngVLA obtains high accuracy (few percent) in the inner disk, and modest accuracy (10\%), even in the outer, fainter regions of the disk. 

The ngVLA has the sensitivity and resolution to trace disk dynamics to larger radii than currently possible -- critical in the search for dark matter in early galaxies. Further, the ngVLA is required to image disk sub-structure, such as spiral arms, rings, bars, or clumps. Such structural elements are the key to understanding the dynamical mechanisms driving star formation in galaxies during the peak cosmic star formation epoch.

\vskip 0.2in
\noindent {\bf References}

Bouwens et al. 2022, in press

Carilli and Walter 2013, ARAA, 51, 105

Carilli et al. 2017, ApJ, 848, 49 

Erb 2006 ApJ 647, 128

van Eymeren et al. 2009, A\& A, 493, 511

Genzel et al. 2020 902, 98

Ginsburg et al. 2015 ASPC 499, 363

Gonzales-Lopez et al. 2019, ApJ, 882, 139

Hunter et al. 2001 ApJ 556, 773

J\'{o}zsa 2007 A\&A 468 ,903

Neeleman et al. 2021 ApJ 911, 141

Riechers et al. 2022, Nature 602, 58

Valentino et al. 2018, ApJ, 869, 27

Walter et al. AJ, 2008, 136, 256

Walter et al. 2016, ApJ, 833, 67

\end{document}